\documentclass{aa}
\usepackage{graphicx}
\usepackage{txfonts}
\usepackage[figuresright]{rotating}
\begin{document}
   \title{The {\itshape XMM-Newton}/{\itshape
   Chandra} monitoring campaign of the Galactic center region}

   \subtitle{Description of the program and preliminary results}

   \author{R. Wijnands,
          \inst{1}
          J.J.M. in 't Zand,
	 \inst{2}
          M. Rupen,
          \inst{3, 4}
	  T. Maccarone,
          \inst{5}
          J. Homan,
          \inst{6}
          R. Cornelisse,
          \inst{5}
	  R. Fender,
	  \inst{5, 1}
	  J. Grindlay,
          \inst{7}
	  M. van der Klis,
	  \inst{1}
          E. Kuulkers,
          \inst{8}
          C.B. Markwardt,
          \inst{4, 9}
          J.C.A. Miller-Jones,
          \inst{1}
	  Q.D. Wang,
          \inst{10}
          }

   \offprints{R. Wijnands}

   \titlerunning{{\itshape XMM-Newton/Chandra} monitoring campaign}
   \authorrunning{Wijnands et al.}

   \institute{Astronomical Institute ``Anton Pannekoek'', University of
   Amsterdam, Kruislaan 403, 1098 SJ, The Netherlands\\
              \email{rudy@science.uva.nl; michiel@science.uva.nl; jmiller@science.uva.nl}
        \and
            SRON National Institute for Space Research, Sorbonnelaan 2, 3584 CA, Utrecht, The Netherlands\\
            \email{jeanz@sron.nl}
	\and
	    National Radio Astronomy Observatory, 1003 Lopezville Road, Socorro, NM 87801, USA
	\and
	    NASA, Goddard Space Flight Center, Greenbelt, MD 20711, USA\\
            \email{mrupen@milkyway.gsfc.nasa.gov; craigm@milkyway.gsfc.nasa.gov}
         \and
             School of Physics and Astronomy, University of Southampton, Southampton SO17 1BJ, UK\\
             \email{tjm@astro.soton.ac.uk, cornelis@astro.soton.ac.uk, rpf@phys.soton.ac.uk}
	\and
	   Center for Space Research, Massachusetts Institute of
	   Technology, 77 Massachusetts Avenue, Cambridge, MA 02139,
	   USA \\
	   \email{jeroen@space.mit.edu}
	\and
           Harvard-Smithsonian Center for Astrophysics, 60 Garden Street, Cambridge, MA 02138, USA\\
           \email{josh@cfa.harvard.edu}
	\and
	   ISOC, ESA/ESAC, Urb. Villafranca del Castillo, P.O. Box 50727, 28080 Madrid, Spain\\	
           \email{Erik.Kuulkers@esa.int}
	\and
	    Department of Astronomy, University of Maryland, College Park, MD 20742, USA
	\and
            Astronomy Department, University of Massachusetts, Amherst, MA 01003, USA\\
            \email{wqd@astro.umass.edu}
 }


\abstract{We present the first results of our X-ray monitoring
campaign on a 1.7 square degree region centered on Sgr A* using the
satellites {\it XMM-Newton} and {\it Chandra}. The purpose of this
campaign is to monitor the behavior (below 10 keV) of X-ray sources
(both persistent and transient) which are too faint to be detected by
monitoring instruments aboard other satellites currently in orbit
(e.g., {\it Rossi X-ray Timing Explorer}; {\it INTEGRAL}). Our first
monitoring observations (using the HRC-I aboard {\it Chandra}) were
obtained on June 5, 2005. Most of the sources detected could be
identified with foreground sources, such as X-ray active stars. In
addition we detected two persistent X-ray binaries (1E 1743.1--2843;
1A 1742--294), two faint X-ray transients (GRS 1741.9--2853; XMM
J174457--2850.3), as well as a possible new transient source at a
luminosity of a few times $10^{34}$ erg s$^{-1}$.  We report on the
X-ray results on these systems and on the non-detection of the
transients in follow-up radio data using the Very Large Array. We
discuss how our monitoring campaign can help to improve our
understanding of the different types of X-ray transients (i.e., the
very faint ones).

 }

   \maketitle
%

\section{Introduction}

Many X-ray sources (the so-called X-ray transients) exhibit orders of
magnitude variability in their X-ray luminosities. Normally they are
too dim to be detected and they are only discovered when they
experience one of their bright outbursts. The brightest transients can
be identified with Galactic neutron stars and black holes accreting
matter from a companion star (the ``X-ray binaries''). The outbursts
are ascribed to a huge increase in the accretion rate onto the compact
object. Several other types of sources can also manifest themselves as
X-ray transients (e.g. accreting white dwarfs, magnetars, $\gamma$-ray
bursts, flare stars, young stellar objects, active binaries), however,
their peak luminosities are many orders of magnitude lower than those
of the transient X-ray binaries (except for $\gamma$-ray bursts and
bursts from magnetars). One usually refers only to transient X-ray
binaries when talking about X-ray transients.

\subsection{Classifying the X-ray transients \label{subsection:classifying}}

Early satellites were mostly able to detect only the brightest
transients. As the instruments became more sensitive, fainter
transients were found which exhibit qualitatively different behavior
from the brighter systems. Consequently, in this paper we classify
the different X-ray transients based on their maximum observed peak
X-ray luminosities (from here on we will quote 2--10 keV luminosities
unless otherwise noted):

\paragraph{{\bf Bright to very bright X-ray transients:}}
These transients have peak X-ray luminosities of $10^{37-39}$ erg
s$^{-1}$ (e.g., \cite{1997ApJ...491..312C}). Monitoring campaigns with
several satellites (e.g., {\it BeppoSAX}, {\it RXTE}, {\it INTEGRAL};
see, e.g., \cite{2001egru.conf..463I, 1996ApJ...469L..33L,
2001ASPC..251...94S, 2004AstL...30..382R}; see
\cite{2001egru.conf..463I} for an overview of past and current
monitoring campaigns focusing on the Galactic center region) have been
very successful in discovering many such bright X-ray transients and
monitoring their X-ray properties.  Intensive studies of the large
amount of available data have yielded a good understanding of their
behavior and it has been found that a large fraction of them harbor
accreting black holes if they accrete from a companion star with a
mass $\la$1 solar mass, but a neutron star if the companion has a mass
$\ga$10 solar masses. In systems with a low-mass companion (the
low-mass transients), the mass transfer occurs because the donor star
overflows its Roche-lobe. To explain the outbursts, the disk
instability model is the most widely accepted (e.g.,
\cite{2001NewAR..45..449L}). In the massive transients the mass
transfer occurs because of the strong stellar wind of the companion
star or through a decretion disk.  The most common type are the
so-called Be/X-ray transients in which matter is accreted from the
circumstellar decretion disks around rapidly spinning B or sometimes
late O-type stars. The physics behind the irregular outbursts observed
for the bright high-mass transients are not yet fully understood but
it has been suggested that these systems might be Be/X-ray binaries
which have relatively low eccentricities (e.g.,
\cite{2001A&A...377..161O}). 

\paragraph{{\bf Faint X-ray transients:}}  
These transients have peak X-ray luminosities of $10^{36-37}$ erg
s$^{-1}$. Although the faint high-mass transients also contain mostly
neutron-star accretors, their outbursts are markedly different from
their brighter siblings since the faint outbursts occur usually in
series separated by the orbital period while the brighter outbursts do
not repeat every orbit, clearly indicating different physics involved
(\cite{2001A&A...377..161O}). It is thought that these periodic
outbursts occur when an accretor is in a wide eccentric orbit around a
Be star and only at minimum distance from the companion does the
accretor come inside the decretion disk of the Be star where its is
able to accrete matter and thus become X-ray active (e.g,
\cite{2001A&A...377..161O}).  Similarly, several detections of
low-mass faint transients have been made in the past (e.g.,
\cite{1990IAUC.5104....1S, 1994ApJ...425..110P}), but only recently it
was realized that they might form a distinct class from the brighter
low-mass systems (\cite{1999ApL&C..38..297H,
2001egru.conf..463I}). This realization came when many such faint
low-mass systems were detected by the Wide Field Cameras (WFC) aboard
{\it BeppoSAX}.  Several global characteristics of the faint low-mass
transients sets them apart from the brighter systems. First, contrary
to what is observed for the brighter low-mass systems, a very large
fraction of the faint ones contain neutron-star accretors as
determined by the detection of type-I X-ray bursts or millisecond
X-ray pulsations\footnote{It is interesting to note that all but one
of the seven currently known accreting millisecond X-ray pulsars (see
\cite{2004AIPC..714..209W} for a review) are faint X-ray transients.
The one which is not a faint transient (XTE J1751--305; peak X-ray
luminosities of a few times $10^{37}$ erg s$^{-1}$) had a very short
outburst duration (e-folding time of only $\sim$7 days) resulting in a
low time-averaged accretion rate similar to the other accreting
millisecond pulsars.} from them. Second, their Galactic distribution
is different from that of the brighter systems
(\cite{2002A&A...392..885C}), with the faint low-mass transients more
concentrated toward the Galactic center. \cite{2000MNRAS.315L..33K}
argued that the faint low-mass transients are indeed different from
the bright systems and that they are mainly neutron star X-ray
binaries in very compact binaries with orbital periods $<$80
minutes. However, clearly not all systems have such short orbital
period since, e.g., the faint transients and millisecond X-ray pulsars
SAX J1808.4--3658, XTE J1814--338, and IGR J00291+5934 have periods
$>$80 minutes. Also not all systems harbor neutron stars since the
faint transient XTE J1118+480 is a strong black hole candidate
(\cite{2001ApJ...556...42W}).

\paragraph{{\bf Very faint X-ray transients (VFXTs):}} These
transients have peak X-ray luminosities of $10^{34-36}$ erg
s$^{-1}$. Until recently, only limited evidence was available for the
existence of this class because detecting such very faint transients
is challenging due to sensitivity and/or angular resolution limits of
many X-ray instruments.  Despite the difficulties in finding VFXTs,
pointed observations with relatively sensitive X-ray satellites (e.g.,
{\it Granat}; {\it ASCA}) resulted in the detection of several VFXTs
near the Galactic center (e.g., \cite{1994ApJ...425..110P,
1998ApJ...508..854T, 1996PASJ...48..417M}). More recently, a
significant number of systems have been found
(\cite{1999ApJ...525..215S, 2004MNRAS.351...31H, 2003A&A...406..299P,
2005MNRAS.357.1211S, 2003ApJ...598..474M}), thanks to the sensitive
{\it Chandra} and {\it XMM-Newton} X-ray satellites, supporting the
claim (e.g., \cite{2005MNRAS.357.1211S}) that a class of VFXTs exists
in our Galaxy. It is very likely that these VFXTs are accreting
neutron stars and black holes since only one accreting white dwarf has
exhibited outbursts above $10^{34}$ erg s$^{-1}$
(\cite{1985MNRAS.212..917W}). For this reason we use $10^{34}$ erg
s$^{-1}$ as the lower luminosity border for the VFXTs class (see also
\cite{1984MNRAS.210..899V}). In section~\ref{subsection:VFXTs} we
discuss the VFXTs further.\\

Our classification of {\it bright to very bright}, {\it faint}, and
{\it very faint} X-ray transients is somewhat arbitrary and it is
clear that systems straddling these classes do exist. For example, the
neutron star X-ray transient in the globular cluster NGC 6440 was
classified as a faint X-ray transient (\cite{1999A&A...345..100I}) but
later it was found to also exhibit bright X-ray outbursts
(\cite{2003ApJ...598..481K}). Furthermore, the bright neutron star
X-ray transient SAX J1747.0--2853 (e.g., \cite{2004A&A...416..311W})
was seen on several occasions at luminosities of only a few times
$10^{35}$ erg s$^{-1}$ (e.g., \cite{2002ApJ...579..422W}). If the
bright outbursts of this source had been missed because, for example,
no X-ray satellite was pointed in the source direction, then it would
have been misclassified as a VFXT. Nevertheless, in this paper such a
classification will prove quite useful in talking about the different
types of transients.

Recently, another type of accreting neutron star has been identified
in the Galaxy: the so-called 'burst-only sources'
(\cite{2002A&A...392..885C}).  The burst-only sources are a group of
nine objects detected by the {\it BeppoSAX}/WFC when they exhibited a
type-I X-ray burst. No detectable accretion emission around the bursts
could be detected with that instrument, with typical upper limits on
the accretion luminosities of the order of $10^{36}$ erg
s$^{-1}$. Subsequent X-ray observations of these systems using a
variety of satellites revealed that one source is a persistent X-ray
source at very faint luminosities (\cite{intzandetal2005}), two are
faint X-ray transients (e.g., \cite{2002A&A...392..931C}), and one is
a VFXT (e.g., \cite{2004MNRAS.351...31H}). The other five sources
could not yet be classified, although they are all X-ray transients
since follow-up observations with {\it Chandra} could not detect any
persistent accretion luminosities from them
(\cite{2002A&A...392..931C}). Presumably, they experienced their X-ray
bursts during brief accretion outbursts which had peak luminosities
below $\sim$$10^{36}$ erg s$^{-1}$. Therefore, these systems are good
candidates to be classified as VFXTs, although definitive proof has to
come from detecting these systems during such very faint outbursts
(see also \cite{2004NuPhS.132..518C} for a discussion on the
classification of burst-only sources).

\subsection{VFXTs in more detail \label{subsection:VFXTs}}

Currently, little is known about the properties of the VFXTs due to
the low number of systems known and the scarcity of observations
during their outburst episodes. Some of them might be intrinsically
bright transients at large distances; however, many are observed near
the Galactic center, indicating source distances of $\sim$8 kpc and
therefore very low peak intrinsic luminosities. In addition, some
VFXTs might be intrinsically brighter than observed due to inclination
effects (e.g., \cite{2005ApJ...633..228M}), but this can be argued for
only a small fraction of the VFXTs (see the appendix; see also
\cite{kingwijnands2005}). Therefore, we consider it likely that most
VFXTs indeed have very faint intrinsic X-ray luminosities.

The characteristics of the VFXTs (e.g., their spectra, their outburst
light curves, timing properties; e.g., \cite{2005MNRAS.357.1211S,
2005ApJ...622L.113M, 1998ApJ...508..854T}) indicate that they are not
a homogeneous class of sources but that different types of accreting
neutron stars and black holes show themselves as VFXTs. The detection
of slow pulsations in some VFXTs (e.g., \cite{1998ApJ...508..854T})
indicates that at least some VFXTs could have a high mass donor star
since a slow X-ray pulsar is usually associated with high-mass X-ray
binaries. The detection of a significant number of high-mass systems
would be very important since the currently identified (brighter)
systems form likely only a small fraction of the total Galactic
population of high-mass X-ray binaries.  However, it is possible that
some of these systems might in fact not be high-mass X-ray binaries
but instead relatively close-by accreting magnetic white dwarfs (i.e.,
the intermediate polars). From the X-ray data it is difficult to
ascertain the exact nature of these systems and identifications of
their optical or IR counterparts are needed to distinguish the
high-mass X-ray binary systems from the white dwarf systems.

No significant pulsations have so far been detected for the other
VFXTs. More information about the nature of these systems could
be obtained by studying them in optical or infrared. However, none of
the VFXTs has so far been detected at optical or infrared
wavelengths. For the VFXTs in the Galactic center region it was found
that they harbor companions fainter than B2 IV stars
(\cite{2005ApJ...622L.113M}). Therefore, it is likely that a
significant fraction of the VFXTs are neutron stars and black holes
accreting matter at a very low rate from a low-mass companion
star. Moreover, at least one VFXT has exhibited type-I X-ray bursts
(SAX J1828.5--1037; \cite{2002A&A...392..885C, 2004MNRAS.351...31H})
which are usually identified with low-mass X-ray binaries.  Such
low-mass X-ray binaries have very low time-averaged accretion rates,
which could become a challenge for our understanding of their
evolution (\cite{kingwijnands2005}).

To improve on the limited amount of knowledge about the observational
properties of VFXTs, monitoring campaigns are required with
instruments which:

\begin{itemize}

	\item are sensitive enough to detect the very faint X-ray
	luminosities of these systems

	\item have a large field-of-view (FOV) to monitor a large
	region allowing for the discovery and monitoring of many
	systems
	
	\item have rapid data turn-around time to allow fast follow-up
	observations at all wavelengths to study the VFXTs when they
	are still active

	\item have (sub-)arcsecond resolution to limit source
	confusion and to allow for unique determination of the
	counterparts at other wavelengths which will help to establish
	the nature of the systems (e.g., IR observations might help to
	determine the type of companion star).

\end{itemize}

Such a monitoring instrument does not exist and in practice might be
difficult to achieve since, for example, a large FOV and very good
sensitivity are usually mutually exclusive. Fortunately, {\it
XMM-Newton} and {\it Chandra} are excellent to perform such a
monitoring program but only within a limited FOV. Therefore, we have
secured such a program using both satellites for a limited region
close to Sgr A*. This region was chosen for two reasons: there is
a high concentration of stars along the line of sight and most known
VFXTs have been detected in this region (e.g.,
\cite{2005ApJ...622L.113M}).  We will first describe the details of
the program and then present some initial preliminary results.

\section{Description of our program}

\begin{figure}
   \centering
      \includegraphics[width=\textwidth]{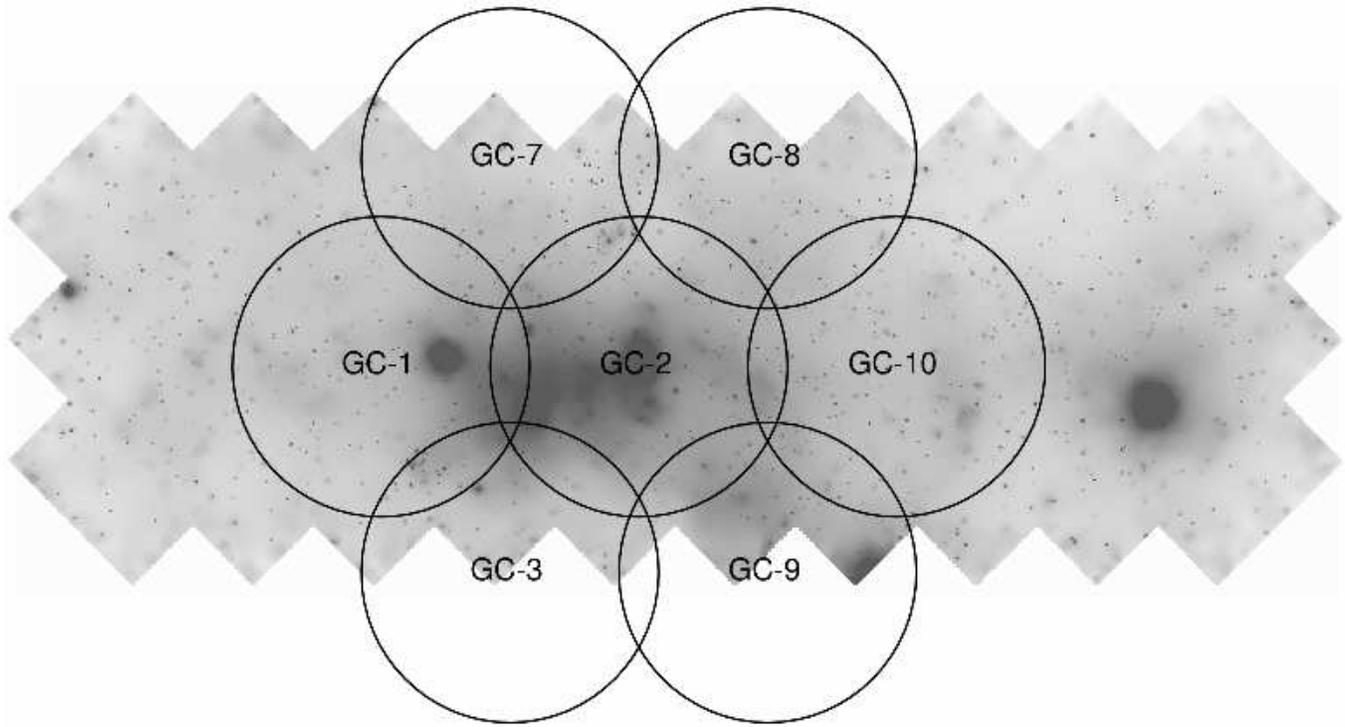}

      \caption{Field-of-view of our monitoring campaign in Galactic
      coordinates. The circles indicate the {\it XMM-Newton}/MOS FOV
      which is similar to the {\it Chandra}/HRC-I FOV. The FOV is
      over-plotted on the {\itshape Chandra} Galactic center survey
      data as presented by \cite{2002Natur.415..148W}, which is
      centered around Sgr A* and covers a region of approximately
      1\degr $\times$ 2\degr. The 'field name' (see
      Tab.~\ref{table:observations}) of each pointing is indicated.}

         \label{fig:fov}
   \end{figure}

In proposal cycle 4 of {\it XMM-Newton}, we have secured a program
using this satellite in combination with the {\it Chandra} X-ray
Observatory to monitor a 1.7 square degree region centered around Sgr
A*. The FOV of our program is shown in Figure~\ref{fig:fov}. We will
have 4 X-ray epochs (two with {\it XMM-Newton} and two with {\it
Chandra}) with each epoch consisting of seven observations (their
pointing directions are indicated in Figure~\ref{fig:fov}).  Each
pointing lasts $\sim$5 ksec and will reach an overall sensitivity of
$\sim$$5\times 10^{33}$ erg s$^{-1}$ (at 8 kpc; from here on we assume
a distance of 8 kpc when quoting X-ray luminosities) for the {\it
Chandra} observations to an order of magnitude better for the {\it
XMM-Newton} observations. The FOVs of the different pointings overlap
by several arcminutes (Fig.~\ref{fig:fov}) to compensate for the loss
in sensitivity toward the edge of the FOV of the instruments.  For the
{\it Chandra} pointings we have chosen to use the HRC-I detector
because it has the largest FOV of any instrument aboard {\it Chandra}
(the FOV of the HRC-I is approximately equal to the FOV of the
instruments aboard {\it XMM-Newton}), although this comes at a loss of
sensitivity for hard sources and the loss of any spectral
information. All detectors will be on during the {\it XMM-Newton}
observations but we will mostly focus on the data obtained with the
EPIC detectors (the two MOS and the pn chips). The RGS will only be
useful when a single bright source is in the FOV.

The purpose of our monitoring campaign is to study the variability
behavior of the transient and persistent sources in the survey region
at levels not reachable by other monitoring instruments in
orbit. Although our interests focus on the accreting neutron stars and
black holes in the FOV, our campaign will also be used to study the
X-ray properties of X-ray active stars, dense star clusters (e.g., the
Arches cluster), flares from Sgr A*, accreting white dwarfs, and any
other object in the FOV which might exhibit persistent or transient
X-ray emission at levels detectable during our observations.  The
different observation epochs are separated in time from each other by
at least one month so that we can monitor the X-ray behavior of the
detected sources on timescales of about 1 month to almost a year. The
first epoch data were gathered on 5 June 2005 ({\it Chandra}) and the
remaining three epochs are currently scheduled for the week of 17--24
October 2005 ({\it Chandra}), and at the end of February 2006 and
early April 2006 ({\it XMM-Newton}). Here we report on the initial
results obtained during the first epoch {\it Chandra} data.

\section{Data analysis}

The first data of our monitoring campaign were obtained on 5 June
2005, using the {\it Chandra} satellite (see
Tab.~\ref{table:observations} for a log of the observations) and the
HRC-I detector.  We processed the data using the {\it Chandra} CIAO
tools (version 3.2.1) and the standard {\it Chandra} analysis
threads\footnote{Available from http://cxc.harvard.edu/ciao/}. We
checked for background flares during our observations, but none were
found, allowing us to use all available data. We merged the 7
different HRC-I observations into one image which we show in
Figure~\ref{fig:images}. Clearly, several bright sources are visible
by eye. We used the tool {\it wavdetect} to search for point sources
in our data and to obtain the coordinates of each source that was
detected. We ran the tool on the combined image as well as on each
individual observation. Due to variations in the size and shape of the
point-spread-function as a function of offset angle from the pointing
directions, we ran {\it wavdetect} on images with different binning
factors. We are still exploring ways to optimize our detection method
so it is likely that in our final analysis we will find more sources
than the ones we report on in this paper. However, we expect that in
the current analysis we are complete for sources with inferred X-ray
luminosities $>$$10^{34}$ erg s$^{-1}$.

The errors on the source positions are difficult to estimate for the
sources found at relatively large offset angles. The asymmetries in
the point-spread-function at large off-axis angles can result in large
systematic uncertainties when using {\it wavdetect} (e.g.,
\cite{hong2005}). We are pursuing extensive simulations to investigate
the effects on the positional errors for a large range of offset
angles as well as different source luminosities. A similar
investigation has already been performed for the {\it Chandra}/ACIS-I
combination by \cite{hong2005}. Although they investigated the ACIS-I
(and not the HRC-I), used only offset angles up to $10'$ (instead of
$>$20$'$ as we sometimes encounter), and focused mainly on the faint
sources, we will use their results as a first order approximation on
the accuracy of the positions we obtain using {\it wavdetect}. We use
equation 5 in \cite{hong2005} to estimate the uncertainties in our
positions. If the sources were detected in multiple observations, we
only give the positions and their errors obtained from the data set in
which the sources had the smallest offset in order to minimize
systematic uncertainties. However, we urge caution when using our
positional uncertainties.

For each detected source we extracted the background corrected count
rate from the images using the standard CIAO tools for the full energy
range of the HRC-I (0.08--10 keV). The count rates obtained can be
converted into fluxes using PIMMS\footnote{Available from
http://cxc.harvard.edu/toolkit/pimms.jsp}, assuming particular
spectral models and values for the interstellar absorption. Again, the
analysis is complicated because many sources are detected at large
off-axis angles and vignetting becomes a serious issue. It is
currently difficult to correct the count rates for vignetting because
the effects depend strongly on both the off-axis angle and the assumed
source spectra (which are unknown since the HRC-I does not currently
allow to extract energy information). Therefore, the count rates we
quote are the uncorrected count rates and are therefore lower limits
to the count rates the sources would have if they were located
on-axis. Depending on the off-axis angles and the source spectra, the
on-axis count rates could have been larger by a factor of a few.
Again, if the sources were detected in multiple observations, we only
give the count rates from the data set in which the sources had the
smallest offset to minimize the systematic errors on the derived
fluxes.

\section{Results}

In total we have detected 21 sources so far. Two sources (the Sgr A*
complex and the Arches cluster) are known to embody a complex of point
sources in combination with strong diffuse emission
(\cite{2003ApJ...589..225M, 2002ApJ...570..665Y}). The analysis of
these complex regions is still in progress. Ten of the remaining
sources can be identified with known stars (e.g., HD 316314, HD
316224, HD 161274, TYC 06840-38-1, ALS 4400; Fig.~\ref{fig:images}
left) or have clear counterparts in the Digital Sky Survey images
indicating that they are foreground objects (and hence have relatively
low X-ray luminosities). We will not discuss the detections of the
foreground objects further in this paper, instead we will focus on the
detected X-ray binaries.

\subsection{The persistent sources}

We detected the two persistent X-ray binaries known to be present in
the surveying region: 1E 1743.1--2843 and 1A 1742--294. 

\subsubsection{1E 1743.1--2843}

1E 1743.1--2843 is a persistent X-ray binary for which the type of
accreting object is not yet known. The source was in the FOV of two of
the seven HRC-I pointings and was detected during both observations
but we detected no bursts from the source.  The position obtained from
our {\it Chandra}/HRC-I data is consistent with that derived from a
previous {\it XMM-Newton} observation (\cite{2003A&A...406..299P})
although, due to the systematic uncertainties in our positional
errors, it cannot currently be determined if our position is better
than the {\it XMM-Newton} one. We used PIMMS to convert the obtained
count rate (see Tab.~\ref{table:binaries}). We assumed an absorbed
power-law model similar to what was found by
\cite{2003A&A...406..299P} when fitting the {\it XMM-Newton}
observation of the source (they obtained an equivalent hydrogen column
density $N_{\rm H}$ of $2\times10^{23}$ cm$^{-2}$ and a photon index
of 1.8). This results in unabsorbed fluxes of $1.8\times 10^{-10}$
(2--10 keV) and $3\times 10^{-10}$ erg cm$^{-2}$ s$^{-1}$ (0.5--10
keV) and X-ray luminosities of 1.4 and $2.3\times 10^{36}$ erg
s$^{-1}$, respectively. These X-ray luminosities are very
similar to what has been seen before for this source (e.g.,
\cite{2003A&A...406..299P}).

\subsubsection{1A 1742--294}

1A 1742--294 is a persistent X-ray binary harboring a neutron-star
accretor as evidenced by the type-I X-ray bursts observed from this
system (see, e.g., \cite{1994ApJ...425..110P}).  We detected this
source during both HRC-I pointings in which the source was in the
FOV. During the GC-10 pointing we detected an X-ray burst. Our {\it
Chandra} position is fully consistent with the best position so far
reported on this source (using {\it ROSAT};
\cite{2001A&A...368..835S}) and despite the possible unknown
systematic uncertainty in our errors, our position is better.  We
again used PIMMS to convert the obtained count rate (see
Tab.~\ref{table:binaries}) and used the absorbed power-law model
($N_{\rm H} \sim$$6\times10^{22}$ cm$^{-2}$ ; photon index $\sim$1.8)
found when fitting the {\it BeppoSAX} and {\it ASCA} data of the
source (\cite{1999ApJ...525..215S,2002ApJS..138...19S}). This results
in unabsorbed fluxes of $2.3\times 10^{-10}$ (2--10 keV) and
$3.7\times 10^{-10}$ erg cm$^{-2}$ (0.5--10 keV) s$^{-1}$. The
corresponding X-ray luminosities are 1.8 and $2.8\times 10^{36}$ erg
s$^{-1}$, consistent with what has been observed before for this
source (e.g., \cite{1999ApJ...525..215S}).

\subsection{The transient sources}

Two transients were clearly visible during our observations: GRS
1741.9--2853 and XMM J174457--2850.3. We made a preliminary
announcement of the detection of these new outbursts on 6 June 2005
(\cite{2005ATel..512....1W}). Following these detections, we obtained
an additional {\it Chandra} observation of both sources (using the
ACIS-I detector) on 1 July 2005 (see Tab.~\ref{table:observations} for
details). Because the two transients were only $\sim$4.6$'$ away from
each other, we could observe both sources with only one ACIS-I
pointing. We placed both sources at an off-axis angle of 7$'$ in order
to limit pile-up in case the sources were as bright as seen during the
HRC-I observations. The ACIS-I data were also analyzed using CIAO and
the standard threads. Again, all data could be used since no episodes
of high background emission occurred during our observation.

\subsubsection{GRS 1741.9--2853}

GRS 1741.9--2853 is a neutron star X-ray transient (it exhibits type-I
bursts; e.g., \cite{1999A&A...346L..45C}) which has been detected
several times in outburst since its original discovery in 1990
(\cite{1990IAUC.5104....1S}). Its peak luminosity is typically a few
times $10^{36}$ erg s$^{-1}$ making it a faint X-ray transient (see
\cite{2003ApJ...598..474M} for more details). This source was detected
during two of our pointings (Tab.~\ref{table:binaries}) but we
detected no bursts. The position of the source was consistent with,
but not better than the one obtained by \cite{2003ApJ...598..474M}.
The observed count rate was converted into fluxes using PIMMS and
assuming an absorbed power-law with $N_{\rm H} = 9.7\times10^{22}$
cm$^{-2}$ and a photon index of 1.88
(\cite{2003ApJ...598..474M}). This results in unabsorbed fluxes of
$1.1\times 10^{-10}$ (2--10 keV) and $1.8\times 10^{-10}$ erg
cm$^{-2}$ s$^{-1}$ (0.5--10 keV), yielding X-ray luminosities of 0.8
and $1.4\times 10^{36}$ erg s$^{-1}$, respectively (for comparison
with previous {\it Chandra} data on this source reported by
\cite{2003ApJ...598..474M}, we also list the 2--8 keV luminosity of
$7.0\times10^{35}$ erg s$^{-1}$). GRS 1741.9--2853 was also detected
during the additional {\it Chandra}/ACIS-I observation
(Fig.~\ref{fig:extra_image}). We extracted the source spectrum using a
source extraction region of 10$''$ and a background extraction circle
of 50$''$ from a source-free region close to GRS 1741.9--2853. The
spectrum was rebinned to have at least 15 counts per bin to allow the
$\chi^2$ fitting method. The resulting spectrum is shown in
Figure~\ref{fig:spectra}. We used XSPEC to fit the spectrum and the
fit results obtained are listed in
Table~\ref{table:spectral_fits}. Clearly, the source flux had
decreased by almost an order of magnitude within about a month (i.e.,
since 5 June 2005). The long-term light curve of the source is plotted
in Figure~\ref{fig:lc} showing the multiple outbursts of the source in
the last 15 years.

\subsubsection{XMM J174457--2850.3}

\begin{figure}
   \centering

\includegraphics[width=0.5\textwidth]{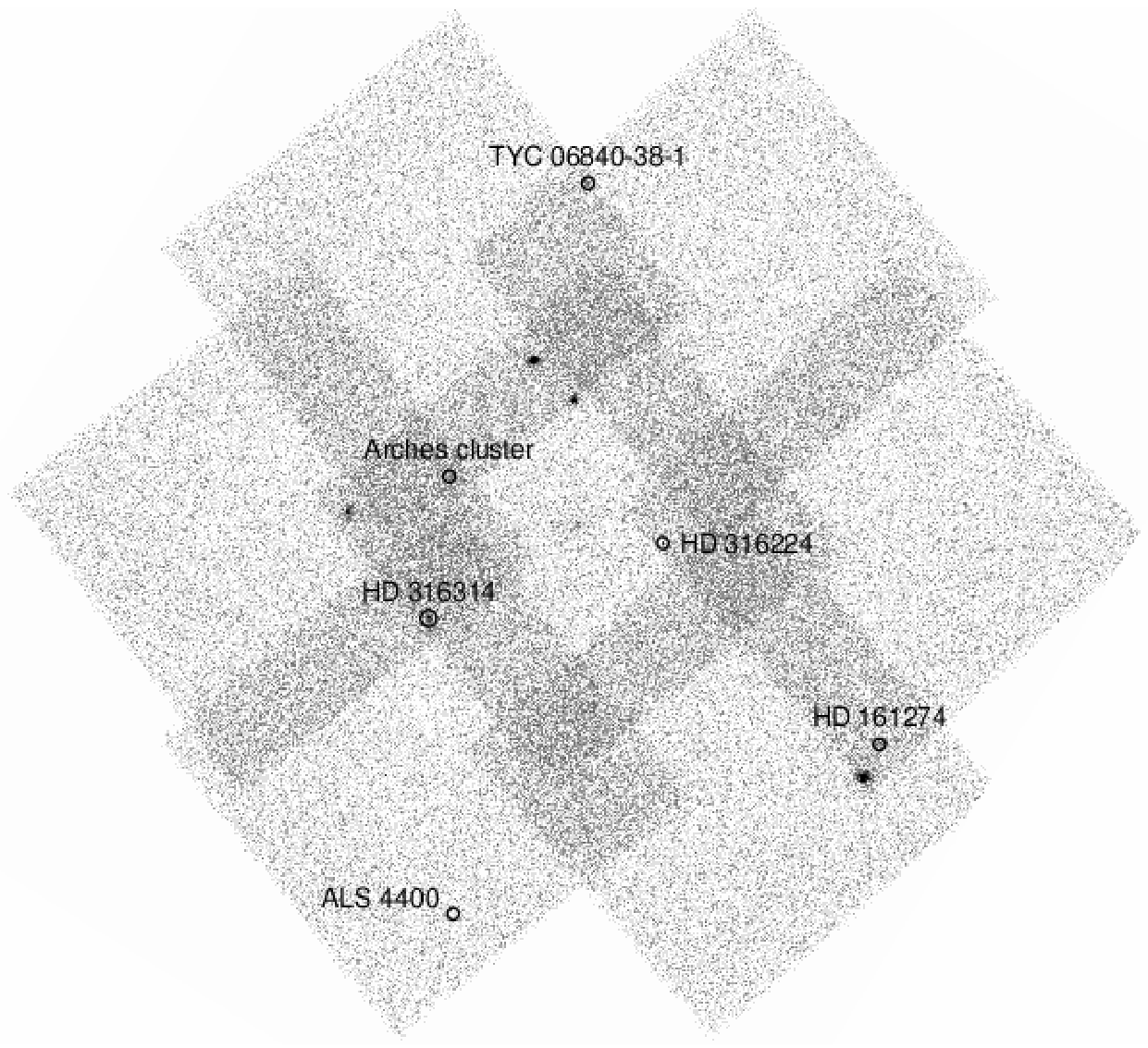}\includegraphics[width=0.5\textwidth]{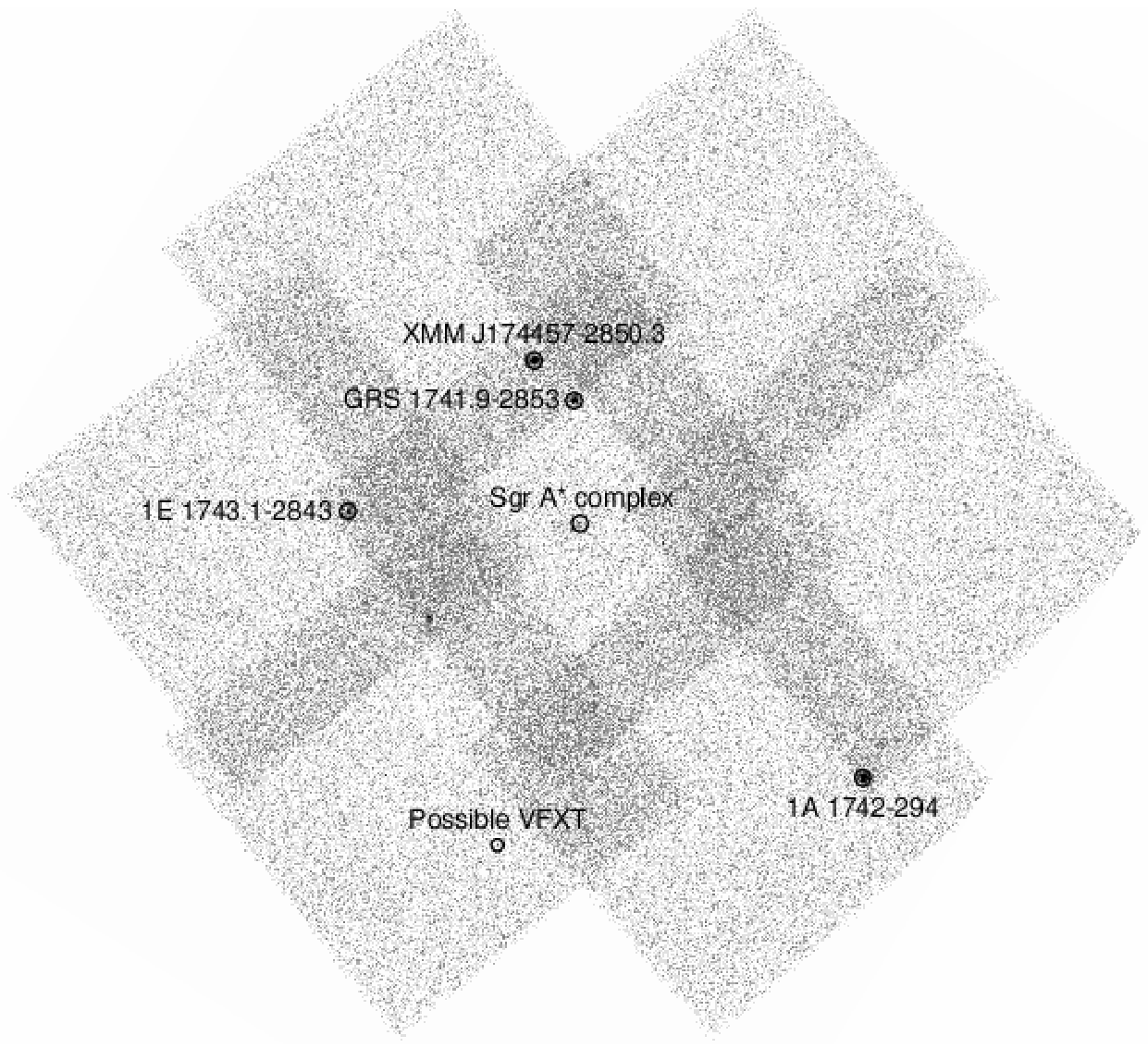}

      \caption{The merged image of the 7 {\itshape Chandra}/HRC-I
      observations. {\bf Left panel:} the Arches cluster and a sub-set
      of the detected foreground objects are indicated (only those
      stars which are known in Simbad are labeled). {\bf Right panel:}
      the detected X-ray binaries as well as the possible VFXT. Also
      indicated is the complex X-ray emission around Sgr A*. }

	\label{fig:images}
      \end{figure}

\begin{figure}
   \centering
	\includegraphics[width=.45\textwidth]{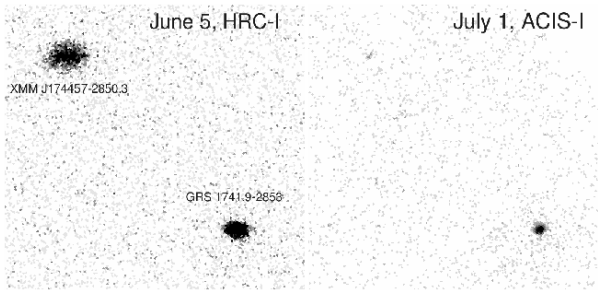}

      \caption{{\bf Left panel:} the {\it Chandra}/HRC-I image of XMM
      J174457--2850.3 and GRS 1741.9--2853 as obtained on June 5, 2005
      (field GC-2), {\bf Right panel:} the {\it Chandra}/ACIS-I image
      of both sources obtained on July 1, 2005. To show the two
      sources most clearly we have rebinned the images to a
      resolution of $\sim$2$''$. The apparent extended nature of the
      sources is due to their relatively large offset positions with
      respect to the pointing direction of the satellite.}

	\label{fig:extra_image}
      \end{figure}

\begin{figure}
   \centering
	\includegraphics[angle=-90,width=0.45\textwidth]{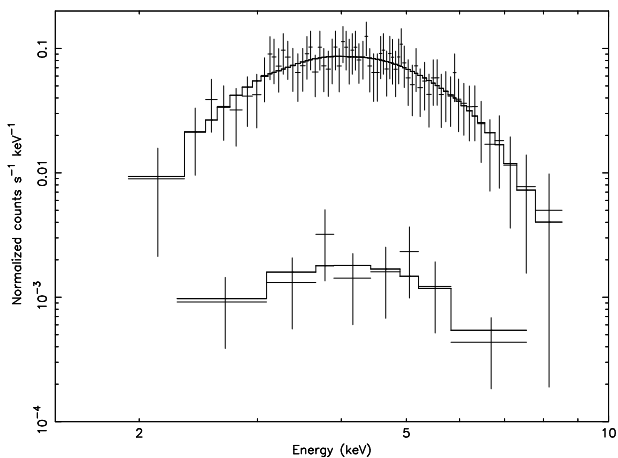}

      \caption{The spectra of GRS 1741.9--2853 (top graph) and XMM
      J174457--2850.3 (bottom graph). The solid lines trough the data
      points indicate the best absorbed power-law fit to the
      data. Before fitting, the data points of GRS 1741.9--2853 were
      rebinned so that each bin has 15 counts. The data of XMM
      J174457--2850.3 were fitted without any rebinning but for
      display purposes that data in the figure are rebinned to 3
      counts per bin.}

	\label{fig:spectra}
      \end{figure}

XMM J174457--2850.3 is also clearly detected during our HRC-I
observations (Fig.~\ref{fig:images}). This source has been detected
only once before in outburst in 2001 (using {\it XMM-Newton};
\cite{2005MNRAS.357.1211S}). During that outburst the source was
seen at a peak luminosity of $5 \times 10^{34}$ erg s$^{-1}$,
justifying a classification as a VFXT. We detected it during two of
our pointings (Tab.~\ref{table:binaries}) but saw no bursts. Our
source position is consistent with that obtained by
\cite{2005MNRAS.357.1211S}, although the exact uncertainty on our
HRC-I position is currently unclear. However, the source was also
detected during our additional ACIS-I observation yielding a more
reliable position even though the source was relatively weak (see
Tab.~\ref{table:binaries}). This position is significantly better than
the {\it XMM-Newton} one.  The observed HRC-I count rate was converted
into fluxes using PIMMS and assuming an absorbed power-law with
$N_{\rm H} = 6\times10^{22}$ cm$^{-2}$ and a photon index of 1.0
(\cite{2005MNRAS.357.1211S}). This resulted in unabsorbed fluxes of
$1.1\times 10^{-10}$ (2--10 keV) and $1.3\times 10^{-10}$ erg
cm$^{-2}$ s$^{-1}$ (0.5--10 keV) and in X-ray luminosities of 0.8 and
$1.0\times 10^{36}$ erg s$^{-1}$, respectively. This is significantly
brighter than what was previously found for the source and makes it a
borderline case as a VFXT. As stated above, XMM J174457--2850.3 was
also detected during the additional {\it Chandra}/ACIS-I observation
(Fig.~\ref{fig:extra_image}). We extracted the source spectrum using a
source extraction region of 5$''$. Due to the rather low number of
source photons (26 counts in the 0.3--7.0 keV energy range) we did not
rebin the spectrum or subtract the background (which was $<$0.3 photon
in the source region and therefore negligible) so that we could use
the Cash statistics (\cite{1979ApJ...228..939C}) when fitting the
spectrum in XSPEC. The fit results obtained for this observation are
also listed in Table~\ref{table:spectral_fits} and the resulting
spectrum is shown in Figure~\ref{fig:spectra}. Clearly, the source
flux had decreased by nearly three orders of magnitude within
approximately a month (i.e., since 5 June 2005). The long-term light
curve of the source is plotted in Figure~\ref{fig:lc}.

\subsubsection{A possible new VFXT}

None of the other known transients in the FOV of our observations (see
Tab.~\ref{table:sources_in_FOV}) were conclusively detected in our
HRC-I data. The upper limits on their luminosities depend strongly on
their spectral shape and their off-axis positions, with a rough
estimate of $\sim$$10^{34}$ erg s$^{-1}$. Several additional weak
sources were detected during our observations which could not be
identified with a star in the Digital Sky Survey database. Only one of
these had a large enough count rate (see Tab.~\ref{table:binaries})
that its X-ray luminosity exceeded $10^{34}$ erg s$^{-1}$ if it had a
'prototypical X-ray binary' spectrum (power-law model with photon
index of 1.8 and a typical $N_{H}$ of $6\times10^{22}$
cm$^{-2}$). Using such a spectral shape, the source had unabsorbed
X-ray fluxes of 1.9 and $3.1\times10^{-12}$ erg s$^{-1}$ for 2--10 keV
and 0.5--10 keV, respectively, and thus luminosities of $1.5\times
10^{34}$ erg s$^{-1}$ (2--10 keV) and $2.4\times10^{34}$ erg s$^{-1}$
(0.5--10 keV).  We note that we do not know the intrinsic source
spectrum and therefore these fluxes and luminosities could be
significantly off if the real source spectrum is considerably
different. We investigated the {\it Chandra} and {\it XMM-Newton}
archives and found that the source was in the FOV of one previous {\it
XMM-Newton} observation. The source was not detected during this {\it
XMM-Newton} observation, but it was at the edge of its FOV making it
difficult to obtain a reliable upper limit on the flux, especially
because we do not know the spectral shape of the source. We estimate
that the luminosity of the source was at least a factor of a few
fainter during the {\it XMM-Newton} observation compared with our
HRC-I data. Although this is suggestive of a transient nature for this
source, it could also be a highly variable persistent
source. Currently, we will refer to this source as a possible new
VFXT.

\subsubsection{Observations at other wavelengths}

We obtained VLA observations at 4 and 6 cm on 8--9 June 2005 of GRS
1741.9--2853, XMM J174457--2850.3, and the possible new VFXT. The
analysis of these radio data is complicated by the strong side-lobes
of Sgr A* and we are still in the process of fully analyzing these
data.  A preliminary analysis of the 4 cm data shows that none of the
sources were conclusively detected with radio fluxes of
$0.003\pm0.060$, $-0.002\pm0.046$, and $0.032\pm0.043$ mJy/beam,
respectively. On 8 June 2005, \cite{2005ATel..522....1L} obtained
I-band images of GRS 1741.9--2853 and XMM J174457--2850.3 using the
Magellan-Baade telescope but could not detect the I-band counterparts
of the sources. This is not surprising when considering the high
absorption column in front of both sources.

\section{Discussion}

We have presented our initial results of the first observations taken
as part of our {\it XMM-Newton}/{\it Chandra} monitoring campaign of
the inner region of our Galaxy. Using our {\it Chandra}/HRC-I
observations we detected mostly foreground objects (like X-ray active
stars), but we also detected two persistent X-ray binaries, two X-ray
transients, and one possible very faint X-ray transient (but its
transient nature requires further confirmation).  Clearly, our
monitoring {\it XMM-Newton}/{\it Chandra} campaign is detecting
transients in outburst which are being missed by the other monitoring
instruments in orbit. Our campaign therefore complements, as designed,
other monitoring campaigns using satellites currently in orbit (e.g.,
\cite{1996ApJ...469L..33L, 2001ASPC..251...94S,
2004AstL...30..382R,2005ATel..438....1K}). These programs find mainly
the brighter transients or the faint transients far away from the
crowded fields near Sgr A*.

The faint X-ray transient GRS 1741.9--2853 was detected at a level of
$\sim$$10^{36}$ erg s$^{-1}$, very similar to what has been observed
previously for this source.  A month after our initial HRC-I
observations this source could still be detected at $\sim$$10^{35}$
erg s$^{-1}$ with the ACIS-I. The parameters obtained for the source
spectrum during this observation were consistent with those found by
\cite{2003ApJ...598..474M} when the source was an order of magnitude
brighter, indicating that the source spectrum is not very dependent on
source luminosity.  Although we did not detect X-ray bursts during our
observations, this source is known to exhibit such phenomena making it
very likely to be a neutron star accreting from a low-mass companion
star. Even though no optical/infrared counterpart has so far been
found for this source, type-I X-ray bursts have only been seen for
low-mass X-ray binaries making it very likely that GRS 1741.9--2853 is
also such a system. The fact that GRS 1741.9--2853 harbors a neutron
star also is consistent with the non-detection of the source in our
radio data since neutron-star low-mass X-ray binaries are known to
exhibit very low radio luminosities (e.g., \cite{2001MNRAS.324..923F,
2005ApJ...626.1020M} )

\begin{figure}
   \centering
	\includegraphics[width=0.45\textwidth]{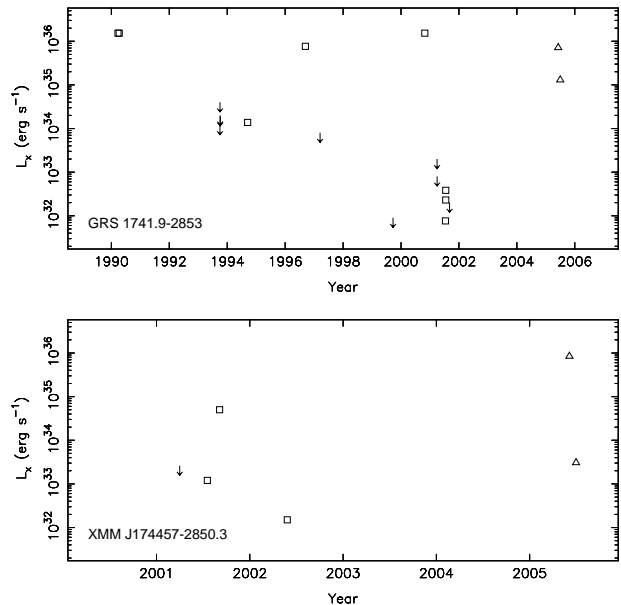}

      \caption{The light curves of GRS 1741.9--2853 (top panel) and
      XMM J174457--2850.3 (bottom panel). In both panels, the
      triangles indicate the new data reported in this paper. In the
      top panel, the squares and the upper limits are taken from
      \cite{2005ApJ...622L.113M} (see this paper for the exact energy
      ranges for each point; our {\it Chandra} luminosities are for
      2--8 keV). In the bottom panel, the squares and the upper limit
      are from \cite{2005MNRAS.357.1211S} and all luminosities are for
      2--10 keV.}

	\label{fig:lc}
      \end{figure}

Figure~\ref{fig:lc} shows that the source has been seen to be in
outburst at least 5 times with X-ray luminosities above $10^{34}$ erg
s$^{-1}$.  Its recurrence time can be estimated to be between 2 and 5
years, making GRS 1741.9--2853 one of the most active transients in
our FOV, with a duty cycle of about 50\% (as estimated from
Fig.~\ref{fig:lc})\footnote{We note that this is likely an upper limit
on the duty cycle since actual source detections are more frequently
reported in the literature than non-detections which will skew the
data toward detections. For example, the data presented by
\cite{2003ApJ...598..474M} of GRS 1741.9--2853 (as used in
Fig.~\ref{fig:lc}) does not report the non-detections of the source as
seen with {\it ROSAT} (\cite{2001A&A...368..835S}) or {\it BeppoSAX}
(\cite{1999ApJ...525..215S}; using the Narrow Field
Instruments). Since no upper limits on the source flux are given in
these papers, we also do not include these non-detections in
Fig.~\ref{fig:lc}. \label{footnote:duty}} .  Its peak luminosity is
very similar to the accreting millisecond X-ray pulsar SAX
J1808.4--3658. For that system and the other accreting millisecond
X-ray pulsars, it has been suggested that their pulsating nature is
related to their rather low time averaged accretion rates (e.g.,
\cite{2001ApJ...557..958C}). Although the time averaged accretion rate
of GRS 1741.9--2853 seems to be higher than for the accreting
millisecond pulsars due to its higher duty cycle, GRS 1741.9--2853
could still be a millisecond X-ray pulsar as well (see also
\cite{2003ApJ...598..474M}), especially if its duty cycle has been
overestimated (see footnote~\ref{footnote:duty}). Unfortunately, its
faintness and its location in the Sgr A* region make it very difficult
to detect these pulsations using {\it RXTE} because of significant
contribution to the detected count rate from other sources in the
FOV. However, with {\it XMM-Newton} pulsations could be detected
within several tens of ksec (depending on the actual fluxes of the
source) if they have similar strengths as the pulsations seen in the
known accreting millisecond pulsars.

The VFXT XMM J174457--2850.3 was also detected during our HRC-I
observations at an X-ray luminosity close to $10^{36}$ erg
s$^{-1}$. This is about a factor of 20 higher than what was previously
seen for this source (\cite{2005MNRAS.357.1211S}). This demonstrates
that VFXTs can exhibit a large range of X-ray luminosities (similar to
what has been observed for the brighter systems) and XMM
J174457--2850.3 is at the border between faint and very faint X-ray
transients, clearly demonstrating that our luminosity boundaries are
somewhat arbitrary as discussed in the introduction.  It is possible
that the previous detection of this source was made either during the
rise or decay of a full outburst and that the maximum luminosity
reached at the time was closer to what we have observed for the source
during our HRC-I observations. Within a month the source luminosity
has decreased by nearly 3 orders of magnitude.  Its X-ray spectrum at
this low X-ray luminosity was consistent with that found by
\cite{2005MNRAS.357.1211S} demonstrating that for this source the
shape of its spectrum is not strongly dependent on luminosity for
luminosities below $5\times 10^{34}$ erg s$^{-1}$. Since we cannot
extract any spectral information from the HRC-I data we cannot
determine if the spectrum was significantly different at times when
the source had X-ray luminosities close to $10^{36}$ erg s$^{-1}$.
Since only very few observations have been performed of this source
(see Fig.~\ref{fig:lc}), it is difficult to estimate its recurrence
time (at most of order 3 years according to Fig.~\ref{fig:lc}) and its
time-averaged accretion rate. The non-detection at radio wavelengths
might indicate that the source harbors a neutron-star accretor since
according to the radio-X-ray correlation found for low luminosity
black hole binaries (\cite{2003MNRAS.344...60G}) the source should
have had (if it harbors a black-hole accretor which was accreting at
$\sim$$10^{36}$ erg s$^{-1}$) a radio flux of $\sim$1 mJy,
significantly higher than our radio upper limit. Alternatively, XMM
J174457--2850.3 could still harbor a black hole, but one which does
not follow this correlation.

We did not detect any unambiguous new VFXTs during our observations,
although we detected a possible new VFXT whose transient nature must
be confirmed. This will be possible with our next sets of monitoring
observations.  Our three additional epochs will also be very important
to find further VFXTs, either previously unknown transients or
recurrent ones. Our observations will allow us to set tighter
constraints on the time averaged accretion rates of these systems than
is currently possible with the available data. Such constraints are
especially important for the low-mass X-ray binaries among the VFXTs
because if their time-averaged accretion rates is very low then our
theories of the evolution of such systems will have a very hard time
explaining their existence without invoking exotic scenarios such as
accretion from a brown dwarf or planet or intermediate mass black hole
accretors (e.g., \cite{kingwijnands2005}). The latter option cannot be
invoked if type-I X-ray bursts have been observed for these systems
since this establishes the existence of a neutron star
accretor. Potential candidates for such systems are the burst-only
sources mentioned in the introduction. Monitoring observations of
these sources with sensitive X-ray telescopes would be very useful to
constrain their time-averaged accretion rates to determine if indeed
these rates are very low for these systems.

Finding new VFXTs and determining their time averaged accretion rate
using our monitoring campaign is only one way forward to increase our
understanding of these enigmatic transients. We now discuss other
avenues that can be explored as well to achieve that goal. First, a
search in the data archives for previously unnoticed VFXTs (e.g., by
comparing different exposures of the same fields) might lead to the
detection of several more systems.  Second, larger regions of our
Galaxy need to be monitored at the desired sensitivity to detect the
low fluxes observed from VFXTs.  It is especially important to
determine if a large number of VFXTs also exist outside the inner
region of our Galaxy. \cite{2005ApJ...622L.113M} found that the excess
of VFXTs within 10 arcminutes of Sgr A* is significant and might point
to an unusual formation history of these systems. However, if a large
number of VFXTs are also found further away from Sgr A* (e.g., systems
like XMM J174716--2810.7 or SAX J1828.5--1037;
\cite{2003ATel..147....1S, 2004MNRAS.351...31H}), then
any production mechanism which requires the high stellar density near
Sgr A* cannot be invoked for these VFXTs.  There is currently no
monitoring satellite in orbit which can perform that task mainly due
to a lack of sensitivity and angular resolution of the
instruments. However, it is possible to derive a first approximation
to the number density of VFXTs at large distances from the center of a
spiral galaxy by performing several deep pointings of the core of
other spiral galaxies. The most obvious choice is the nearest large
spiral galaxy to our own, M31. Within a $\sim$250 ksec exposure it is
possible to observe a 3.7 kpc $\times$ 3.7 kpc region of M31 using the
{\it Chandra}/ACIS-I detector with a limit sensitivity of $1-4 \times
10^{34}$ erg s$^{-1}$ (depending on the spectral properties of the
sources). Several such deep pointings would detect all but the
faintest X-ray transients in a large region of M31. Alternatively,
such programs can also be performed for the smaller spiral galaxy M33
or for galaxies further away. In the latter case, the limiting
sensitivity will be of course less.

\begin{acknowledgements}
RW thanks Michael Muno and Andrew King for useful discussions about
very faint X-ray transients. We also thank Michael Muno for the
information on the absence of eclipses in CXOGC J17535.5--290124.
\end{acknowledgements}


\begin{table*}
\caption{Log of the {\itshape Chandra} observations for epoch 1}       
\label{table:observations}    
\centering                   
\begin{tabular}{l c c c c}      
\hline\hline                 
Field name         & ObsID & Date                   &  Exposure  & Instrument \\
                    &       & (June 5, 2005, UTC)    &  (ksec)    & \\
\hline     
GC-1                & 6188  &  01:44 -- 03:32        &  5.08      & HRC-I \\
GC-2                & 6190  &  03:32 -- 05:05        &  5.15      &  ''     \\
GC-3                & 6192  &  05:05 -- 06:37        &  5.14      &  ''\\
GC-7                & 6194  &  06:37 -- 08:09        &  5.14      &  ''\\
GC-8                & 6196  &  08:09 -- 09:41        &  5.14      &  ''\\
GC-9                & 6198  &  09:41 -- 11:13        &  5.15      &  ''\\
GC-10               & 6200  &  11:13 -- 13:08        &  5.15      &  ''\\
\hline
GRS 1741.9--2853 \& & 6311  &  02:04 -- 03:40 July 1 &  4.01 & ACIS-I\\
XMM J174457--2850.3 &       &                        &      &    '' \\
\hline                                 
\end{tabular}
\end{table*}

\begin{sidewaystable*}
\begin{minipage}[t]{\columnwidth}
\caption{The X-ray binaries detected during our observations}       
\label{table:binaries}    
\centering                   
\renewcommand{\footnoterule}{}  
\begin{tabular}{c c c c c c c l }      
\hline\hline                 
Source name         & In FOV    & Offset\footnote{Offset between the source position and the pointing position} & \multicolumn{3}{c}{Coordinates}  & Count rate\footnote{Count rates are for the full {\itshape Chandra}/HRC-I energy range (0.08--10 keV) or the 0.3--7 keV energy range for the {\itshape Chandra}/ACIS-I; they are background corrected, but are not corrected for offset.}       & Comment  \\
                     & of frame & ($'$)  & RA    &  Dec  & Error ($''$)\footnote{The errors on the coordinates are calculated using equation 5 in \cite{hong2005} with the addition of a 0.7$''$ pointing uncertainty, corresponding to 95\% confidence levels.}    & (counts s$^{-1}$)&         \\
\hline     
1E 1743.1--2843     & GC-1     & 6.6    & 17 46 21.094 & -28 43 42.3 &  1.1            & 0.247$\pm$0.007  & Persistent X-ray binary                   \\
                    & GC-2     & 18.5   &              &             &                 &                  &                                           \\
1A 1742--294        & GC-9     & 11.1   & 17 46 05.201 & -29 30 53.3 &  1.3            & 0.84$\pm$0.01    & Persistent neutron star low-mass X-ray binary        \\   
                    & GC-10    & 20.6   &              &             &                 &                  & Burst detected during GC-10               \\                              
GRS 1741.9--2853    & GC-2     & 9.9    & 17 45 02.385 & -28 54 50.2 &  1.5            & 0.267$\pm$0.008  & Neutron star  faint X-ray transient   \\
                    & GC-7     & 15.9   &              &             &                 &                  &                                           \\
                    & ACIS-I   & 6.9    & 17 45 02.350 & -28 54 49.9 &  1.2            & 0.269$\pm$0.008  &                                           \\
XMM J174457--2850.3 & GC-7     & 11.4   & 17 44 57.440 & -28 50 20.3 &  1.5            & 0.337$\pm$0.009  & Very faint X-ray transient                \\
                    & GC-2     & 13.6   &              &             &                 &                  &                                           \\
                    & ACIS-I   & 7.0    & 17 44 57.451 & -28 50 21.1 &  2.1            & 0.0064$\pm$0.0015&                                           \\
\hline
Possible VFXT       & GC-3     & 8.7    & 17 47 37.671 & -29 08 09.6 &  2.5            & 0.007$\pm$0.002  &  Transient nature needs confirmation       \\
\hline
\end{tabular}
\end{minipage}
\end{sidewaystable*}

\begin{table*}
\begin{minipage}[t]{1.1\columnwidth}
\caption{Spectral results for GRS 1741.9--2854 and XMM J174457--2850.3}       
\label{table:spectral_fits}    
\centering              
\renewcommand{\footnoterule}{}  
\begin{tabular}{l c c}      
\hline\hline                 
Parameter                                            &  GRS 1741.9--2854    & XMM J174457--2850.3         \\
\hline     
$N_{\rm H}$ ($10^{22}$ cm$^{-2}$)                    & $10.5^{+4.9}_{-3.7}$ & 6\footnote{Parameter fixed to the value found by \cite{2005MNRAS.357.1211S}.} \\
Photon index                                      & $1.8^{+1.0}_{-0.8}$  & $1.3\pm1.1$                 \\
Flux ($10^{-12}$ erg cm$^{-2}$ s$^{-1}$, unabsorbed) & & \\                           
~~~~ 0.5--10.0 keV                                   & $37^{+93}_{-14}$     & $0.5^{+0.4}_{-0.5}$         \\
~~~~ 2.0--10.0 keV                                   & $22^{+10}_{-3}$      & $0.4^{+0.2}_{-0.4}$         \\
~~~~ 2.0--8.0 keV                                    & $17^{+12}_{-3}$      & $0.3^{+0.1}_{-0.3}$         \\
\hline                                 
\end{tabular}
\end{minipage}
\end{table*}

\newpage\clearpage

\newpage\clearpage

\appendix

\section{Distribution of transients in the FOV \label{section:appendix}}

\begin{figure}
   \centering
      \includegraphics[width=0.45\textwidth]{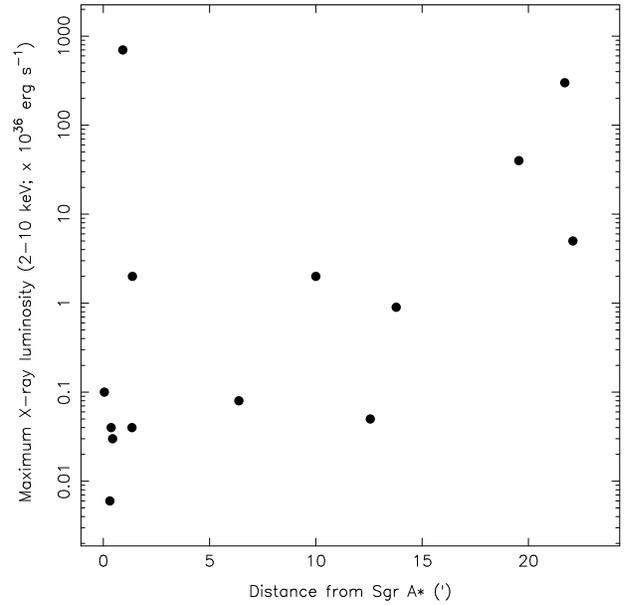}

      \caption{The observed maximum X-ray luminosities (2--10 keV) of
      the X-ray transients in our FOV as a function of the distance of
      the transients to Sgr A*. \label{fig:Lx_v_distance} }

\end{figure}

In Table~\ref{table:sources_in_FOV} we have listed the known X-ray
transients and persistent X-ray binaries which are in the FOV of our
{\it XMM-Newton} and {\it Chandra} monitoring observations. This table
also shows the distance of the sources with respect to Sgr A* and the
maximum reported X-ray luminosity for these sources as found in the
literature. We have converted the luminosities to the 2--10 keV energy
range using the reported spectral parameters of the sources. From this
table and from Figure~\ref{fig:Lx_v_distance} it can clearly be seen
that only one bright to very bright transient is present (1A
1742--289) within a distance of $<$$15'$ from Sgr A*, but eight
VFXTs\footnote{The VFXT CXOGC J174535.5--290124 is located in the
error circle of the faint transient and eclipsing source AX
J1745.6--2901. It is possible that both sources are the same one,
which would reduce the number of VFXTs to seven although this would
not affect the conclusions in this appendix. However, no eclipses were
found for CXOGC J174535.5--290124 in the {\it Chandra} data available
for this source (M. Muno 2005, private communication), making it less
likely that both sources are the same one.} and two faint X-ray
transients (although the two faint transients are just barely brighter
than $10^{36}$ erg s$^{-1}$ and they could just be the brightest VFXTs
in the FOV). Clearly, the number of VFXTs within 15$'$ of Sgr A* is
significantly larger than that of the brighter transients. When
extending the region out to 25$'$, we see eight VFXTs, three faint
transients and only 3 bright to very bright transients. Note, however,
that the region $>$10$'$ away from Sgr A* has been less sampled with
sensitive X-ray instruments than closer to Sgr A*, meaning that the
number of VFXTs within $25'$ of Sgr A* is likely to grow thanks to our
monitoring program.

This strongly suggests that the number density of VFXTs is indeed
significantly higher than that of the brighter transients. The fact
that the brighter systems were until recently much easier to detect
than the fainter systems means that this discrepancy in number
densities will only become larger in the future. However, one VFXT
(CXOGC J174540.0--290031) and one faint transient (AX J1745.6--2901)
have exhibited eclipses and there is strong evidence
(\cite{2005ApJ...633..228M}) that at least CXOGC J174540.0--290031 was
intrinsically much brighter ($>$10$^{36}$ erg s$^{-1}$) than what we
observe because the inner part of the system is blocked from our line
of sight and we only observe the scattered (e.g., in a corona) X-ray
emission from the system, making it artificially seem very faint. One
can argue that this holds true for the other eclipsing source and
possibly for all VFXTs for which we have not yet observed the eclipses
or the X-ray dips associated with high inclination. If we
conservatively assume that sources with inclinations in the range of
60$^\circ$--90$^\circ$ could appear as VFXTs, then a random
distribution of orbital inclinations should give roughly equal numbers
of bright transients and VFXTs (see also
\cite{kingwijnands2005}; the solid angle is proportional to the cosine
of the inclination). Clearly, the distribution of observed transients
in the FOV of our monitoring campaign (see
Tab.~\ref{table:sources_in_FOV}) shows a lack of brighter systems,
even when also considering the two persistent X-ray binaries.

Moreover, strong evidence exists that for the known dippers and
eclipsing X-ray binaries in other parts of the Galaxy we {\it do}
directly look at the inner part of the systems and not just indirectly
via scattering (e.g., due to the detection of kHz QPOs or nearly
coherent oscillations during type-I X-ray bursts in several high
inclination sources; e.g., \cite{2000A&A...361..121B,
2000ApJ...539..847H, 2001ApJ...549L..71W,
2001ApJ...549L..85G}). Therefore, the observed X-ray luminosity is
indeed the intrinsic luminosity of these systems. Only for the
so-called accretion-disk-corona sources (which have the highest
inclination of all systems) do we have evidence that only the
scattered emission is seen making these systems appear fainter than
they are intrinsically. The inclination range for such sources is
significantly smaller than what we used above, thus making the problem
even worse. Clearly, it is very likely that most of the VFXTs do not
appear very faint as a result of inclination effects but rather that
they are intrinsically very faint.

The fact that VFXTs seem to be overabundant close to the Galactic
center compared to brighter systems might also have consequences for
certain types of models for the VFXTs. For example,
\cite{kingwijnands2005} discussed briefly the possibility that the
VFXTs harbor neutron stars and black holes which accrete from the weak
wind of low-mass companion stars resulting in very faint
outbursts. However, they argued that these systems must eventually
evolve into brighter states because the companion stars will fill
their Roche lobes at a certain time in the future. If indeed, as they
suggested, these brighter states have longer durations than the wind
accreting states, then the lack of brighter systems present among the
very faint ones would suggest that this model cannot explain the
nature of the VFXTs.

\begin{table*}
\begin{minipage}[t]{1.0\columnwidth}
\caption{Persistent and transients X-ray binaries in the FOV of our observations}       
\label{table:sources_in_FOV}    
\centering                   
\begin{tabular}{l c l l c c}     
\hline\hline                 
Source name & Distance from Sgr A* & Maximum luminosity\footnote{These
maximum observed luminosities are taken from the references and converted to a 2--10 keV luminosity for a distance of 8 kpc}
& Classification & Comment & Reference\footnote{References: 1: \cite{2005ApJ...622L.113M}; 2: \cite{1976MNRAS.175P..47B}; 3: \cite{1996PASJ...48..417M}; 4: \cite{2003ApJ...598..474M}; 5: \cite{2005MNRAS.357.1211S}; 6: this paper; 7: \cite{2004A&A...416..311W}; 8: \cite{1996ApJ...469L..25G}; 9: \cite{1991AdSpR..11..187I}; 10: \cite{1999A&A...345..826C}; 11: \cite{2002ApJS..138...19S} } \\ & ($'$) & (erg s$^{-1}$) & &
& \\
\hline    
\multicolumn{5}{c}{{\bf Transient sources}}\\ 
CXOGC J174540.0-290031 &  0.05   &     $1\times 10^{35}$    &    very faint & eclipser, radio counterpart     & 1 \\
CXOGC J174541.0-290014 &  0.31   &     $6\times 10^{33}$    &    very faint &  & 1\\ 
CXOGC J174540.0-290005 &  0.37   &     $4\times 10^{34}$    &    very faint &    & 1\\
CXOGC J174538.0-290022 &  0.44   &     $3\times 10^{34}$    &    very faint &    & 1 \\
1A 1742-289            &  0.92   &     $7\times 10^{38}$      &    very bright & radio counterpart          & 2\\
CXOGC J174535.5-290124\footnote{This source is located in the error circle of AX J1745.6--2901 and it is possible that both are the same source. However, no eclipses were found for CXOGC J174535.5--290124 in the {\it Chandra} data available for this source (M. Muno 2005, private communication) making this identification less likely.}               &   1.35   &     $4\times 10^{34}$    &    very faint &    & 1\\
AX J1745.6-2901        &  1.37   &     $2\times 10^{36}$      &    faint & burster, eclipser  & 3\\
CXOGC J174554.3-285454 &  6.38   &     $8\times 10^{34}$      &    very faint   &  & 1 \\
GRS 1741.9-2853        & 10.00   &     $2\times 10^{36}$      &    faint & burster	         & 4\\
XMM J174544-2913.0     & 12.56   &     $5\times 10^{34}$      &    very faint &    &  5\\
XMM J174457-2850.3     & 13.78   &     $9\times 10^{35}$      &    very faint   &             & 6 \\
SAX J1747.0-2853       & 19.55   &     $4\times 10^{37}$      &    bright & burster           & 7\\
GRO J1744-28           & 21.71   &     $3\times 10^{38}$    &    bright & X-ray pulsar      & 8\\
KS 1741-293            & 22.09   &     $5\times 10^{36}$    &    faint	&                 & 9\\
\hline
\multicolumn{5}{c}{{\bf Persistent sources}}\\
1E 1743.1-2843       &18.99      &    $3\times 10^{36}$      &      faint   &  & 10\\
1A 1742-294           &30.95      &    $1\times 10^{37}$      &     bright & burster  & 11\\
\hline                                 
\end{tabular}
\end{minipage}
\end{table*}

\begin{thebibliography}{}

\bibitem[Boirin et al.~(2000)]{2000A&A...361..121B} Boirin, L., Barret, D., 
Olive, J.~F., Bloser, P.~F., \& Grindlay, J.~E.\ 2000, \aap, 361, 121 
 

\bibitem[Branduardi et al.~(1976)]{1976MNRAS.175P..47B} Branduardi, G., 
Ives, J.~C., Sanford, P.~W., Brinkman, A.~C., \& Maraschi, L.\ 1976, 
\mnras, 175, 47P 

\bibitem[Cash (1979)]{1979ApJ...228..939C} Cash, W.\ 1979, \apj, 228, 939 
 
\bibitem[Chen et al.~(1997)]{1997ApJ...491..312C} Chen, W., Shrader, C.~R., 
\& Livio, M.\ 1997, \apj, 491, 312 
 

\bibitem[Cocchi et al.~(1999)]{1999A&A...346L..45C} Cocchi, M., Bazzano, A., 
Natalucci, L., Ubertini, P., Heise, J., Muller, J.~M., \& in 't Zand, 
J.~J.~M.\ 1999, \aap, 346, L45 
 

\bibitem[Cornelisse et al.~(2002a)]{2002A&A...392..885C} Cornelisse, R., et 
al.\ 2002a, \aap, 392, 885 
 
\bibitem[Cornelisse et al.~(2002b)]{2002A&A...392..931C} Cornelisse, R., 
Verbunt, F., in't Zand, J.~J.~M., Kuulkers, E., \& Heise, J.\ 2002b, \aap, 
392, 931 
 
\bibitem[Cornelisse et al.~(2004)]{2004NuPhS.132..518C} Cornelisse, R., et 
al.\ 2004, Nuclear Physics B Proceedings Supplements, 132, 518 
 



\bibitem[Cremonesi et al.~(1999)]{1999A&A...345..826C} Cremonesi, D.~I., 
Mereghetti, S., Sidoli, L., \& Israel, G.~L.\ 1999, \aap, 345, 826 
 


\bibitem[Cumming et al.~(2001)]{2001ApJ...557..958C} Cumming, A., Zweibel, 
E., \& Bildsten, L.\ 2001, \apj, 557, 958 
 

\bibitem[Fender \& Kuulkers (2001)]{2001MNRAS.324..923F} Fender, R.~P., \& 
Kuulkers, E.\ 2001, \mnras, 324, 923 
 

\bibitem[Gallo et al.~(2003)]{2003MNRAS.344...60G} Gallo, E., Fender, R.~P., 
\& Pooley, G.~G.\ 2003, \mnras, 344, 60 
 

\bibitem[Galloway et al.(2001)]{2001ApJ...549L..85G} Galloway, D.~K., 
Chakrabarty, D., Muno, M.~P., \& Savov, P.\ 2001, \apjl, 549, L85 
 

\bibitem[Giles et al.~(1996)]{1996ApJ...469L..25G} Giles, A.~B., Swank, 
J.~H., Jahoda, K., Zhang, W., Strohmayer, T., Stark, M.~J., \& Morgan, 
E.~H.\ 1996, \apjl, 469, L25 
 
\bibitem[Hands et al.~(2004)]{2004MNRAS.351...31H} Hands, A.~D.~P., Warwick, 
R.~S., Watson, M.~G., \& Helfand, D.~J.\ 2004, \mnras, 351, 31 
 



\bibitem[Heise et al.~(1999)]{1999ApL&C..38..297H} Heise, J., in't Zand, 
M.~J.~J., Smith, S.~M.~J., Muller, M.~J., Ubertini, P., Bazzano, A., 
Cocchi, M., \& Natalucci, L.\ 1999, Astrophysical Letters Communications, 
38, 297 

\bibitem[Homan \& van der Klis (2000)]{2000ApJ...539..847H} Homan, J., \& 
van der Klis, M.\ 2000, \apj, 539, 847 
 

\bibitem[Hong et al.~(2005)]{hong2005} Hong, J., van den Berg, M., Schlegel, 
E.M., Grindlay, J.E., Koenig, X., Laycock, S., Zhao, P. 2005, ApJ, in press


\bibitem[In 't Zand et al.~(1991)]{1991AdSpR..11..187I} in 't Zand, 
J.~J.~M., et al.\ 1991, Advances in Space Research, 11, 187 

\bibitem[In 't Zand et al.~(1999)]{1999A&A...345..100I} in 't Zand, 
J.~J.~M., et al.\ 1999, \aap, 345, 100 
 

\bibitem[In't Zand (2001)]{2001egru.conf..463I} in't Zand, J.\ 2001, ESA 
SP-459: Exploring the Gamma-Ray Universe, 463 
 

\bibitem[In 't Zand et al.~(2005)]{intzandetal2005}In 't Zand, J.J.M., Cornelisse, R., 
\& Mendez, M., 2005, A\&A, in press

\bibitem[Kaaret et al.~(2003)]{2003ApJ...598..481K} Kaaret, P., Zand, 
J.~J.~M.~i., Heise, J., \& Tomsick, J.~A.\ 2003, \apj, 598, 481 
 

\bibitem[King (2000)]{2000MNRAS.315L..33K} King, A.~R.\ 2000, \mnras, 315, 
L33 

\bibitem[King \& Wijnands (2005)]{kingwijnands2005} King, A.R. \& 
Wijnands, R. 2005, \mnras Letters, in press (astro-ph/0511486)

\bibitem[Kuulkers et al.~(2005)]{2005ATel..438....1K} Kuulkers, E., et al.\ 
2005, The Astronomer's Telegram, 438, 1 



\bibitem[Lasota (2001)]{2001NewAR..45..449L} Lasota, J.-P.\ 2001, New 
Astronomy Review, 45, 449 


\bibitem[Laycock et al.~(2005)]{2005ATel..522....1L} Laycock, S., Zhao, P., 
Torres, M.~A.~P., Wijnands, R., Steeghs, D., Grindlay, J., Hong, J., \& 
Jonker, P.~G.\ 2005, The Astronomer's Telegram, 522, 1 
 

\bibitem[Levine et al.~(1996)]{1996ApJ...469L..33L} Levine, A.~M.,
 Bradt, H., Cui, W., Jernigan, J.~G., Morgan, E.~H., Remillard, R.,
 Shirey, R.~E.,
\& Smith, D.~A.\ 1996, \apjl, 469, L33 
 


\bibitem[Maeda et al.~(1996)]{1996PASJ...48..417M} Maeda, Y., Koyama, K., 
Sakano, M., Takeshima, T., \& Yamauchi, S.\ 1996, \pasj, 48, 417 


\bibitem[Muno et al.~(2003a)]{2003ApJ...589..225M} Muno, M.~P., et al.\ 2003a, 
\apj, 589, 225 
 

\bibitem[Muno et al.~(2003b)]{2003ApJ...598..474M} Muno, M.~P., Baganoff, 
F.~K., \& Arabadjis, J.~S.\ 2003b, \apj, 598, 474 
 


\bibitem[Muno et al.~(2005a)]{2005ApJ...622L.113M} Muno, M.~P., Pfahl, E., 
Baganoff, F.~K., Brandt, W.~N., Ghez, A., Lu, J., \& Morris, M.~R.\ 2005a, 
\apjl, 622, L113 

\bibitem[Muno et al.~(2005b)]{2005ApJ...626.1020M} Muno, M.~P., Belloni, T., 
Dhawan, V., Morgan, E.~H., Remillard, R.~A., \& Rupen, M.~P.\ 2005b, \apj, 
626, 1020 
 
\bibitem[Muno et al.~(2005)]{2005ApJ...633..228M} Muno, M.~P., Lu, J.~R., 
Baganoff, F.~K., Brandt, W.~N., Garmire, G.~P., Ghez, A.~M., Hornstein, 
S.~D., \& Morris, M.~R.\ 2005, \apj, 633, 228 

\bibitem[Negueruela et al.~(2005)]{negueruela}Negueruela, I., Smith, D.~M., 
Reig, P., Chaty, S., Miguel Torrejon, J. 2005, To appear in
Proceedings of "The X-ray Universe 2005", held in San Lorenzo de El
Escorial (Madrid, Spain), 26-30 September 2005, ESA-SP 604
(astro-ph/0511088)


\bibitem[Okazaki \& Negueruela (2001)]{2001A&A...377..161O} Okazaki, A.~T., 
\& Negueruela, I.\ 2001, A\&A, 377, 161 
 

\bibitem[Pavlinsky et al.~(1994)]{1994ApJ...425..110P} Pavlinsky, M.~N., 
Grebenev, S.~A., \& Sunyaev, R.~A.\ 1994, \apj, 425, 110
 

 \bibitem[Porquet et al.~(2003)]{2003A&A...406..299P} Porquet, D., Rodriguez, 
J., Corbel, S., Goldoni, P., Warwick, R.~S., Goldwurm, A., \& Decourchelle, 
A.\ 2003, \aap, 406, 299 

\bibitem[Porquet et al.~(2005)]{2005A&A...430L...9P} Porquet, D., Grosso, 
N., Burwitz, V., Andronov, I.~L., Aschenbach, B., Predehl, P., \& Warwick, 
R.~S.\ 2005, \aap, 430, L9 
 
\bibitem[Revnivtsev et al.~(2004)]{2004AstL...30..382R} Revnivtsev, M.~G., 
et al.\ 2004, Astronomy Letters, 30, 382 

 


\bibitem[Sakano et al.~(2002)]{2002ApJS..138...19S} Sakano, M., Koyama, K., 
Murakami, H., Maeda, Y., \& Yamauchi, S.\ 2002, \apjs, 138, 19 
 


\bibitem[Sakano et al.~(2005)]{2005MNRAS.357.1211S} Sakano, M., Warwick, 
R.~S., Decourchelle, A., \& Wang, Q.~D.\ 2005, \mnras, 357, 1211 
 
\bibitem[Sidoli et al.~(1999)]{1999ApJ...525..215S} Sidoli, L., Mereghetti, 
S., Israel, G.~L., Chiappetti, L., Treves, A., \& Orlandini, M.\ 1999, 
\apj, 525, 215 


\bibitem[Sidoli et al.~(2001)]{2001A&A...368..835S} Sidoli, L., Belloni, T., 
\& Mereghetti, S.\ 2001, \aap, 368, 835 
 

\bibitem[Sidoli \& Mereghetti (2003)]{2003ATel..147....1S} Sidoli, L., \& 
Mereghetti, S.\ 2003, The Astronomer's Telegram, 147, 1 
 
\bibitem[Sunyaev (1990)]{1990IAUC.5104....1S} Sunyaev, R.\ 1990, \iaucirc, 
5104, 1 
 


\bibitem[Swank \& Markwardt (2001)]{2001ASPC..251...94S} Swank, J., \& 
Markwardt, C.\ 2001, ASP Conf.~Ser.~251: New Century of X-ray Astronomy, 
251, 94 



\bibitem[Torii et al.~(1998)]{1998ApJ...508..854T} Torii, K., et al.\ 1998, 
\apj, 508, 854 
 

\bibitem[Verbunt et al.~(1984)]{1984MNRAS.210..899V} Verbunt, F., Elson, R., 
\& van Paradijs, J.\ 1984, \mnras, 210, 899 
 
\bibitem[Wagner et al.~(2001)]{2001ApJ...556...42W} Wagner, R.~M., Foltz, 
C.~B., Shahbaz, T., Casares, J., Charles, P.~A., Starrfield, S.~G., \& 
Hewett, P.\ 2001, \apj, 556, 42 
 

\bibitem[Wang et al.~(2002)]{2002Natur.415..148W} Wang, Q.~D., Gotthelf, 
E.~V., \& Lang, C.~C.\ 2002, \nat, 415, 148 
 

\bibitem[Watson et al.~(1985)]{1985MNRAS.212..917W} Watson, M.~G., King, 
A.~R., \& Osborne, J.\ 1985, \mnras, 212, 917 
 

\bibitem[Werner et al.~(2004)]{2004A&A...416..311W} Werner, N., et al.\ 
2004, \aap, 416, 311 
 

\bibitem[Wijnands (2004)]{2004AIPC..714..209W} Wijnands, R.\ 2004, AIP 
Conf.~Proc.~714: X-ray Timing 2003: Rossi and Beyond, 714, 209 


\bibitem[Wijnands et al.~(2001)]{2001ApJ...549L..71W} Wijnands, R., 
Strohmayer, T., \& Franco, L.~M.\ 2001, \apjl, 549, L71 
 


\bibitem[Wijnands et al.~(2002)]{2002ApJ...579..422W} Wijnands, R., Miller, 
J.~M., \& Wang, Q.~D.\ 2002, \apj, 579, 422 
 

\bibitem[Wijnands et al.~(2005)]{2005ATel..512....1W} Wijnands, R., et al.\ 
2005, The Astronomer's Telegram, 512, 1 

 

\bibitem[Yusef-Zadeh et al.(2002)]{2002ApJ...570..665Y} Yusef-Zadeh, F., 
Law, C., Wardle, M., Wang, Q.~D., Fruscione, A., Lang, C.~C., \& Cotera, 
A.\ 2002, \apj, 570, 665 
 

\end{thebibliography}
\end{document}